\documentclass[journal]{IEEEtran}
\usepackage[table]{xcolor}
\usepackage{psfrag}
\usepackage{acronym}
\usepackage{etoolbox}
\usepackage{wrapfig}
\usepackage{epsfig}
\usepackage{colortbl,color}
\usepackage{amsmath}
\usepackage{amssymb}
\usepackage{mathabx}
\usepackage{enumerate}
\usepackage{bbm}
\usepackage{algorithm}
\usepackage{algorithmic}
\usepackage{lipsum,amsmath,multicol}
\usepackage{stfloats}
\usepackage{arydshln} 
\usepackage{amsfonts}
\usepackage{dsfont}
\usepackage{upgreek}
\usepackage{pgfplots} 
\usepackage{soul}
\usepackage{tabularx}
\usepackage{multirow}
\usepackage{graphicx}
\usepackage{pgfplots}

\usepackage{cite}
\usepackage{verbatim}

\usepackage{tabularx}
\usepackage{enumitem, ragged2e}
\usepackage{smartdiagram}
\usesmartdiagramlibrary{additions}
\makeatletter
\newcommand*{\compress}{\@minipagetrue}
\makeatother
\acrodef{5G}{fifth generation}
\acrodef{6G}{sixth generation}
\acrodef{AI}{artificial intelligence}
\acrodef{UAV-I}{{\em UAV intelligence}}
\acrodef{AI/ML}{artificial intelligence/machine learning}
\acrodef{CNPC}{control and non-payload communication}
\acrodef{A2A}{agent-to-agent}
\acrodef{AWGN}{additive white Gaussian noise}
\acrodef{BS}{base station}
\acrodef{CRLB}{Cram\'er-Rao Lower Bound}
\acrodef{CDF}{cumulative density function}
\acrodef{CDR}{correct detection rate}
\acrodef{CIR}{channel impulse response}
\acrodef{CRAS}{connected robotics and autonomous systems}
\acrodef{DEC-POMDP}{decentralized partially observable Markov decision process}
\acrodef{D2D}{device-to-device}
\acrodef{EKF}{extended Kalman filter}
\acrodef{EIRP}{effective isotropic radiated power}
\acrodef{KF}{Kalman filter}
\acrodef{FAR}{false alarm rate}
\acrodef{FIM}{Fisher Information matrix}
\acrodef{FL}{federated learning}
\acrodef{FMCW}{frequency modulated continuous wave}
\acrodef{GLRT}{generalized likelihood ratio test}
\acrodef{GP}{Gaussian process}
\acrodef{GPI}{generalized policy iteration}
\acrodef{IoT}{Internet-of-Things}
\acrodef{IS}{image similarity}
\acrodef{KPI}{key performance indicator}
\acrodef{LLRT}{log-likelihood ratio test}
\acrodef{LOS}{line-of-sight}
 \acrodef{MAC}{medium access control}
\acrodef{MAP}{maximum a-posteriori probability}
\acrodef{MEC}{multi-access edge computing}
\acrodef{mMTC}{massive machine type communication}
\acrodef{eMBB}{enhanced mobile broadband}

\acrodef{MMSE}{minimum mean squared error }
\acrodef{MARL}{multi-agent reinforcement learning}
\acrodef{MDP}{Markov decision process}
\acrodef{MIMO}{multiple input multiple output}
\acrodef{ML}{machine learning}
\acrodef{MLE}{maximum likelihood estimator}
\acrodef{mm-wave}{millimeter-wave}
\acrodef{NOMA}{non-orthogonal multiple access}
\acrodef{NLOS}{non line-of-sight}
\acrodef{OG}{occupancy grid}
\acrodef{PF}{particle filtering}
\acrodef{pdf}{probability density function}
\acrodef{PFA}{probability of false alarm}
\acrodef{PL}{packet loss}
\acrodef{POMDP}{partially observable Markov decision process}
\acrodef{PER}{packet error rate}
\acrodef{RCS}{radar cross section}
\acrodef{RMSE}{root mean square error}
\acrodef{RFID}{radiofrequency identification}
\acrodef{RL}{reinforcement learning}
\acrodef{ROC}{receiver operating characteristics}
\acrodef{RR}{reading range}
\acrodef{RRCS}{root-radar cross section}
\acrodef{RV}{random variable}
\acrodef{SLAM}{simultaneous localization and mapping}
\acrodef{SNR}{signal-to-noise ratio}
\acrodef{SIR}{sequential importance resampling} 
\acrodef{TD}{temporal-difference}
\acrodef{THz}{Terahertz}
\acrodef{UAV}{unmanned aerial vehicle}
\acrodef{U2U}{UAV-to-UAV}
\acrodef{UKF}{Unscented Kalman Filter}
\acrodef{URLLC}{ultrareliable and low-latency communication}
\acrodef{VLC}{visible light communication}

\definecolor{aliceblue}{rgb}{0.94, 0.97, 1.0}
\definecolor{greenyellow}{rgb}{0.7, 0.9, 0.4}
\definecolor{caribbeangreen}{rgb}{0.0, 0.8, 0.6}
\definecolor{celadon}{rgb}{0.67, 0.88, 0.69}
\definecolor{lightpastelpurple}{rgb}{0.89, 0.81, 0.98}
\definecolor{lightgreen}{rgb}{0.9, 0.96, 0.9}
\pgfplotsset{compat=1.17}
\begin{document}

\title{Networks of  {UAVs}  of Low--Complexity for \\ Time--Critical Localization} % Applications}

\author{Anna~Guerra,~\IEEEmembership{Member,~IEEE,}
Francesco~Guidi,~\IEEEmembership{Member,~IEEE,}
       \\ Davide~Dardari,~\IEEEmembership{Senior,~IEEE,}
        and~Petar~M.~Djuri\'c,~\IEEEmembership{Fellow,~IEEE}% <-this % stops a space
\thanks{A.~Guerra and D.~Dardari are with the WiLAB - Department of Electrical and Information Engineering ``Guglielmo Marconi" - CNIT, University of Bologna, Italy. E-mail: (anna.guerra3, davide.dardari)@unibo.it. F. Guidi is with WiLAB, CNR-IEIIT, Italy. E-mail: francesco.guidi@ieiit.cnr.it.}% <-this % stops a space
\thanks{P.~M.~Djuri\'c is with the Electrical and Computer Engineering Department, Stony Brook University, Stony Brook, NY 11794, USA.
E-mail: petar.djuric@stonybrook.edu.}% <-this % stops a space
}

%\markboth{Submitted to IEEE Aerospace \& Electronic Systems Magazine}%
%{Guerra \MakeLowercase{\textit{et al.}}: Bare Demo of IEEEtran.cls for IEEE Journals}



\maketitle


\begin{abstract}
%\textcolor{blue}{Alternative title: Networks of Low-Complexity {UAVs}  for Time-Critical Localization-based Applications}
Future networks of \acp{UAV} will be tasked to carry out ever--increasing complex operations that are time--critical and that require accurate localization performance (e.g., tracking the state of a malicious user). Since there is the need to preserve low \ac{UAV} complexity while tackling the challenging goals of missions in effective ways, one key aspect is %represented by 
the \ac{UAV-I}. {The UAV's intelligence includes} %, here intended as
the {\ac{UAV}'s} capability to  process  information and to  make decisions, e.g., to decide where to sense {and whether} to delegate some  tasks  to  other  network  entities. In this paper, we provide an overview of possible solutions for the design of \acp{UAV} of low complexity, %the UAV's intelligence, %\ac{UAV-I}, 
showing some of the needs of the \acp{UAV} for running efficient localization operations, performed either as a team or individually. {Further, we focus} on different network configurations, {which possibly include assistance with edge computing}. %, % or not,
We also discuss open problems and future perspectives for these settings.
\end{abstract}

\begin{IEEEkeywords}
Unmanned aerial vehicles, Localization, Policy learning, Inference engine.
\end{IEEEkeywords}

%\textcolor{red}{Regular articles: Normally, submitted manuscripts are of a length to fit in 6-10 published pages. It should be engaging, and of a technical depth appropriate to be read by a typical AESS member. Figures must be of high quality and should be in color. Bibliographies should generally not exceed 20 referenced items. Brief abstracts are published as a group in the "In this Issue Technically" section. Author photos and bios are not published}

\bstctlcite{IEEEexample:BSTcontrol}

% \fi
\acresetall
\IEEEpeerreviewmaketitle
\section{Motivations}
\label{sec:motivations}

{Robots could be simpler and operate longer if their computational operations, e.g., their ``brains", were delegated to external networks.} %{\color{red} DD: I would remove "with lower latency" as it seems in contrast with that said in the following.}
But it is still an open research problem how much intelligence has to be delegated to external entities when time–critical tasks have to be accomplished. Indeed, \acp{UAV}, like robots, usually operate in uncommon situations, including natural disasters or search–and–rescue applications, where they are needed to react quickly, i.e., with the shortest possible delays. There are two major reasons why time plays a critical role for networks of \acp{UAV}: \textit{(i)} the technology: \acp{UAV} can fly for a limited amount of time because of their {limited battery life}; \textit{(ii)} the considered application: \acp{UAV} can be employed in search–and–rescue applications where targets (e.g., victims of a natural disaster) must be detected and localized as quickly as possible. 
{Another example} of challenging application is when a dense swarm, or even several swarms, of low-cost, energy constrained, and small \acp{UAV} {are flown}  over cities for detecting and tracking non-authorized ({\em malicious}) \acp{UAV} \cite{WanEtAl:J21,GuvEtAl:J18}. {The malicious \acp{UAV} can compromise} %harm 
the security of the population and can be barely detected using conventional terrestrial radars. All of these examples {represent} time–critical localization-based applications. Indeed, with this term, we indicate all the situations when a real-time fast optimal deployment of \acp{UAV} is vitally important for the successful completion of the mission itself, as in %such as for 
disaster relief {settings}. %networks. 
%\textcolor{red}{%If ever longer autonomy will be enabled by hardware advancements,
{Mission-critic applications %, such as search--and--rescue operations,
impose strict limited mission times that should be accounted for when designing the \ac{UAV} control  \cite{guerra2020access}.}
%} %{\color{red} DD: I don't understand this statement. I would have expected the contrary. } 

In all such events, preserving low latency and low complexity for transmission and data processing becomes an essential requirement to limit the mission time and to make decisions quickly. In particular, the delays of each phase of the \ac{UAV} mission should be optimized, i.e., during sensing, information processing, navigation, communication with peers or with cellular infrastructure.
For example, {peer-to-peer communications, involving shorter links between \acp{UAV}, are favorable to reduce energy consumption. {However, in large swarms, it could be difficult to guarantee a full connectivity, especially when the communication range is shorter} 
\cite{zeng2019accessing}.} 

A possible solution is to partially or entirely offload some of the computations to other entities belonging to higher layers of the communication infrastructure, e.g., to the edges. This allows for keeping the complexity of the UAVs low, thus enabling network scalability. {On the other hand}, it presents  challenging demands on the design of the \ac{UAV} control \cite{guerra2020dynamic,GuvEtAl:J18,guerra2020access}.
In fact, edge-aided solutions ask for a re-design of the communication and collaboration architecture between the \acp{UAV}. %To this end, %it is reassuring that the \ac{6G} of wireless systems promises the integration of \ac{UAV} networks within the cellular infrastructure
In this direction, preliminary studies foresee that the \ac{6G} wireless networks will integrate the \ac{UAV}-based flying networks within the cellular infrastructure, facilitating the connection with clouds or edges with high reliability and lower latency \cite{SaaBenChe:J19,popovski2018wireless}. %{\color{red} DD: I would say "In this direction, preliminary studies foresee that \ac{6G} wireless networks will integrate \ac{UAV}-based flying networks within the cellular infrastructure ...." }

In this paper, we study a network of {autonomous \acp{UAV} with low-complexity} that performs three main functional tasks: sensing, learning, and communication. The \acp{UAV} observe some features of the environment ({\em sensing}), then estimate their states (e.g., the \ac{UAV} positions or some features of the surrounding) and learn which actions should be taken through interactions and experience ({\em learning}). In this sense, we refer to the \ac{UAV-I} as the ability to process the information, to collaborate, and to navigate, i.e., to learn a control policy that can maximize some mission-related performance.  %Instead, 
With {\em sensing} we indicate the operation that allows \acp{UAV} to collect information from the environment. {Sensing can be active, where each \ac{UAV} interrogates the environment with probing signals, or passive, where the sensors measure some physical parameter.} 
%It can be either active or passive whether the \acp{UAV} are used to interrogate the environment with probing signals or they are only in reception. 
All the collected data are then processed for inferring some useful features for the considered application. For example, the \acp{UAV} estimate the positions of the targets by collecting range and angle measurements through on-board low--cost radars \cite{guerra2020dynamic}, or they optimize their positions for reducing the risk of electromagnetic field exposure \cite{lou2021green}. The main contributions {of this paper} can be summarized as follows:
\begin{itemize}
    \item We discuss about {\em how} and {\em when} it is better to distribute the on–board \ac{UAV-I} over high-layer entities of the network. This discussion is also corroborated by an example of complexity analysis; %while performing target tracking;
    \item We overview architectures and techniques that can be used by \acp{UAV} for improving the information processing, the communication scheme, and autonomous navigation in unknown environments, which can be critical in collaborative {applications}, especially when localization and tracking tasks have to be accomplished, for example, in search--and--rescue scenarios;
    \item We provide two examples of case study to corroborate our analysis. We first address a target tracking application with a model--based information seeking control. Then, we consider a mapping and target detection application where the \acp{UAV} learn the best trajectory through trials--and--errors over episodes (model-free navigation). In both cases, time {for completing the tasks is limited.} %is considered a limited resource.
\end{itemize}

\begin{figure*}[t!]
 \centering
 \includegraphics[width=0.89\linewidth,draft=false]
 {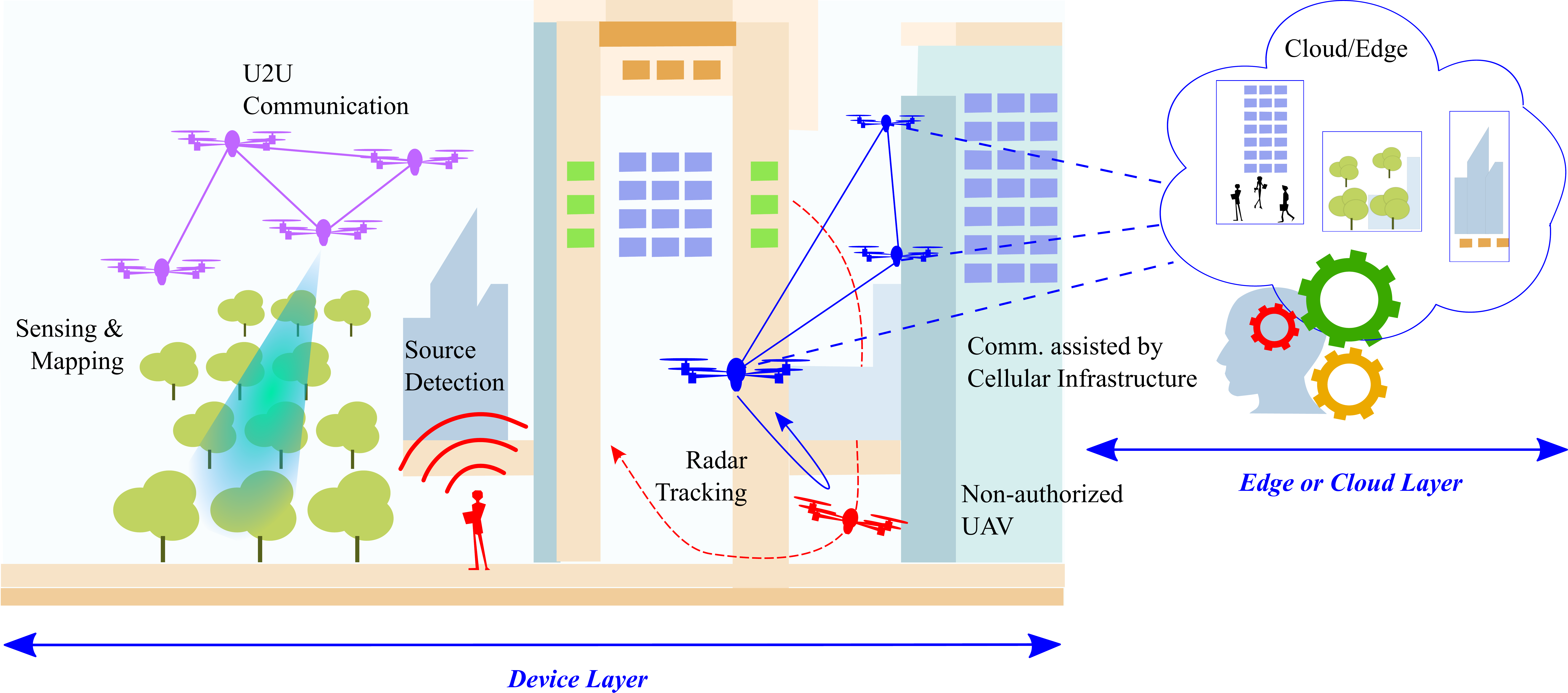}
 \caption{The considered networks of low-complexity UAVs.}
 \label{fig:fig1}
 \end{figure*}

The rest of the paper is organized as follows. Section~\ref{sec:network} describes the network of \acp{UAV} with low--complexity in terms of architectures and communication strategies. Then, the \ac{UAV-I} {for state inference} is introduced in %Sec.~\ref{sec:intelligence}
Sec.~\ref{sec:inference} and policy evaluation {is presented in} Sec.~\ref{sec:policy}. Finally,  case studies are discussed in Sec.~\ref{sec:example} and conclusions are drawn in Sec.~\ref{sec:future}.

\section{Networks of Low-Complexity UAVs}
\label{sec:network}
\begin{figure*}[t!]
 \centering
 \includegraphics[width=0.89\linewidth,draft=false]
 {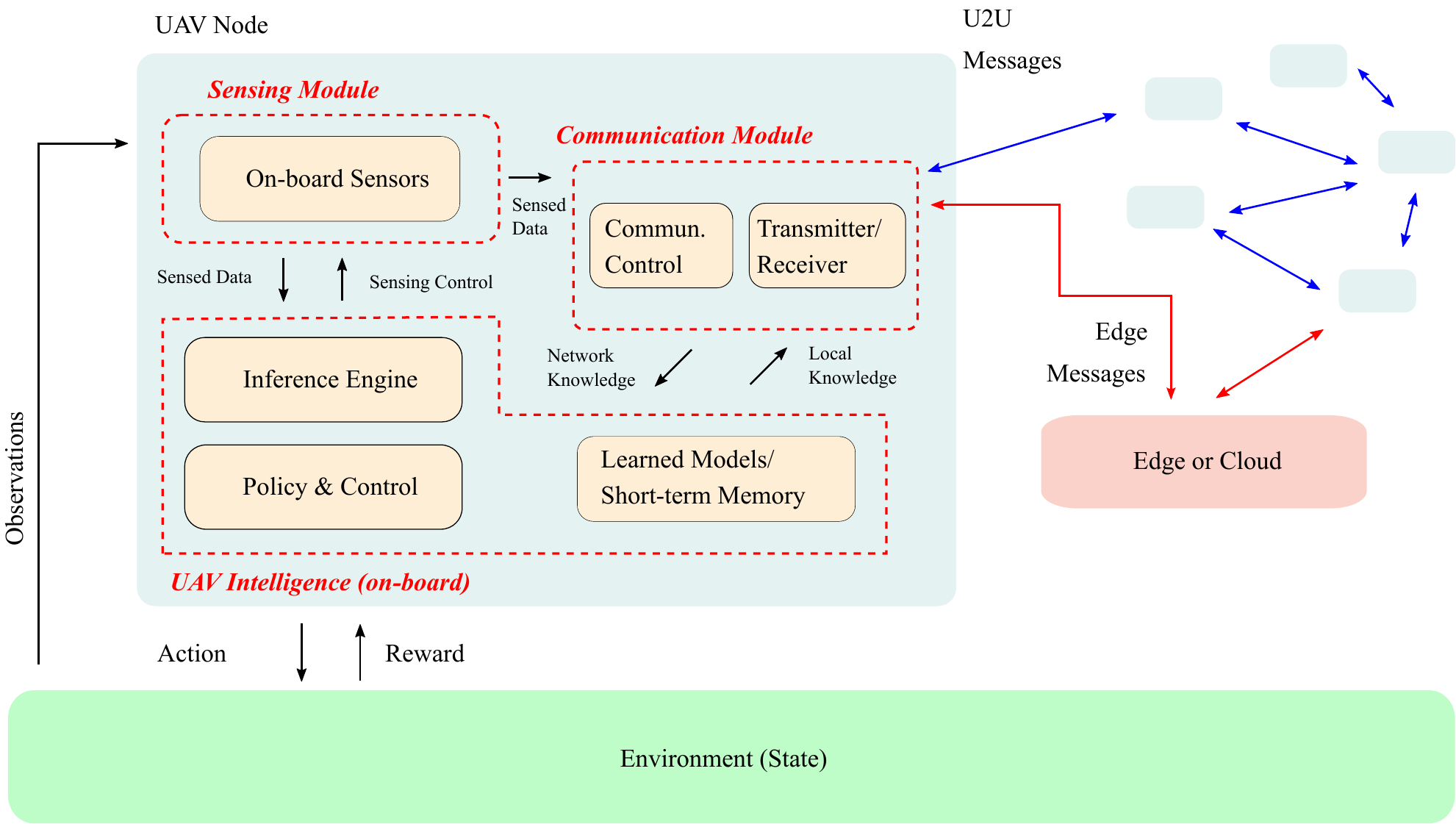}
 \caption{The overall architecture of a UAV agent.}
 \label{fig:fig2}
 \end{figure*}

In Fig.~\ref{fig:fig1}, {we display} a couple of examples of localization-based applications. On the left, there is a swarm {of \acp{UAV}} dedicated to sensing and environmental monitoring in remote areas, i.e., areas difficult to reach by ground vehicles and hardly covered by a cellular network. {Conversely, on the right, the swarm operates} in an urban area {to detect and track a non-authorized \ac{UAV}}, and it is assisted by an infrastructure. %To describe these two cases, the main network and communication architectures for \acp{UAV} will be detailed below.
The network of  \acp{UAV} with low-complexity, in charge of the mission, is depicted in Fig.~\ref{fig:fig2}. There we see the main modules of a \ac{UAV} (also referred to as an agent in the sequel) %are shown in Fig.~\ref{fig:fig2}. %and its interactions with the environment, with other \acp{UAV}, and with the cellular network. More specifically, 
%three main functional blocks can be 
with the following functionalities: %envisioned as follows: 
{\em (i) a Sensing Module} consisting of on--board sensors {designed
for observing the environment}; %through which the environment can be observed; 
{\em(ii) a \ac{UAV-I} Module} where the sensed data are processed for estimating the state of the environment and for controlling the \ac{UAV} movements and communications; {\em (iii) a Communication Module} for regulating the communications with peers or with the cellular network.

%In this section, we discuss how the network architecture can help the \ac{UAV-I} distribution over the communication network. 
\ac{UAV-I} {refers to} the capability of a \ac{UAV} to process information and to make decisions concerning sensing or delegating some tasks to other network entities. 
%In particular, we review different \ac{UAV} network architecture, either peer-to-peer (blue arrows in Fig.~\ref{fig:fig2}) or assisted by the cellular infrastructure (red arrows in Fig.~\ref{fig:fig2}), and their respective communication protocols. 
%\subsection{UAV Network Architectures \& Communications}
%\label{sec:arch}
%As pointed out earlier, when networks are comprised of low-cost \acp{UAV}, their lack of processing capabilities may require offloading some of the computations to higher layers of the network, e.g., to the cloud or edge nodes, as in Fig.~\ref{fig:fig2}. However, delegating too much \ac{UAV-I} to external entities can reduce the capability of the \acp{UAV} to autonomously perform the assigned tasks, especially in time-critical applications, where a fast reaction is preferable over higher accuracies. 
%
Note also that the \ac{UAV-I} distribution over the network impacts the choice of the communication strategy to connect either with a peer \ac{UAV} or with an edge. In fact, {\em communication for} networks of \acp{UAV} can rely either on {\em ad-hoc} \ac{D2D} protocols or it can be assisted by cellular networks \cite{popovski2018wireless}. Moreover, it can entail the exchange of sensors/measurement data ({\em payload communication}) %{\color{red} DD:why mission-related data? are these sensors/measurement data?  } 
or \ac{CNPC} messages (e.g., sense-and-avoid commands) \cite{zeng2019accessing}. Payload data, like those coming from sensors, require higher data rates and longer latency to be transmitted because of their informative contents and length, while \ac{CNPC} needs higher reliability because of its safety-critical contents \cite{zeng2019accessing}.
%Thus, the choice of the \ac{UAV-I} distribution affects also the communication scheme and the entailed latency.

%In the following we discuss about two possible architectures that depends where and how learning is performed.
%\ac{U2U} and infrastructure-based network and data exchanges. 
In the following, we discuss {the impact of } \ac{D2D} (i.e., \ac{U2U}) and infrastructure-based {communication schemes on \ac{UAV} networks}. Notably, this choice does not prevent from designing networks where the two solutions are alternated depending on the needs of the considered application.
%It is important to remark that this choice does not prevent from designing networks where the two solutions are alternated depending on the needs of the considered application.

\subsection{\ac{U2U} Network}
In this kind of network, each \ac{UAV} can be considered as a simple and low-complexity agent that can acquire a local knowledge of the environment through observations collected by its sensors and processing engines. When the \ac{UAV-I} is fully on-\ac{UAV} and not delegated to external entities, the learning task is performed locally, possibly by exchanging data with other \acp{UAV}. Thus, each \ac{UAV} works with a limited set of data collected by its sensors and \ac{U2U} exchanges (i.e., messages with sensed or processed data and \ac{UAV} positioning information).

When working with scarce and limited information (e.g., isolated clusters of \acp{UAV}), the inference may not be accurate and the computational complexity might become an issue, especially for low-cost \acp{UAV}. On the other hand, decisions can be made much faster, and the network itself can promptly be reconfigured while becoming more fault-tolerant.

% With \ac{U2U} communication, we refer to a 
In this network scenario, data are directly exchanged among \acp{UAV} without routing them towards an infrastructure. Two possible solutions are foreseen \cite{popovski2018wireless}: {\textit{(i)}} {\em ad-hoc} \ac{U2U} communication that can be either in-band, or out-of-band if they exploit unlicensed bands; 
\textit{(ii)} \ac{U2U} communication with the {supervision} of 5G and beyond cellular infrastructure. Cellular network generations 4G LTE Release-12 and 5G already support \ac{D2D} communication, enabling the connection between physically close devices through a side-link  \cite{popovski2018wireless}.

In the first solution, \acp{UAV} are not supervised and they can form clusters simply considering the shortest distance among them (e.g., via multi-hops) or according to a social-aware framework that allows them to trust each other (e.g., through advanced social graphs) \cite{saxena2020exploiting}. 
In this case, the \acp{UAV} are fully responsible for setting up, controlling and coordinating the communications. This solution is important for critical applications, as in natural disasters, where the cellular infrastructure may be temporarily collapsed.
In the second configuration, instead, the \acp{BS} are used for communication network management and, also, as trusted authorities to create links between the \acp{UAV}. 

Notably, in both configurations, \ac{U2U} communication helps in improving spectral efficiency by offloading the cellular network. This becomes of tremendous importance when a massive number of {non-\ac{UAV}} users is simultaneously sharing the cellular infrastructure, as expected in beyond \ac{5G} and \ac{6G}. Moreover, it allows for improving the flexibility and {for reducing} the latency, when a {one}-hop is possible, and packet compression schemes are implemented so that the payload size is reduced.

{In networks with a large number of \acp{UAV}}, coverage may require multi-hop communications with the risk of the information becoming aged. %, and consequently useless, due to increased end-to-end delay.
In this case, to preserve low latency, some packets may be dropped and, thus, never delivered. As a result, this limits the availability of information at each \ac{UAV}, which in turn might impact the accuracy of the processing and decision-making. Finally, fully \ac{U2U} requires that the \ac{UAV-I} is not delegated to external entities, with all the consequences they pertain to the hardware complexity, cost, and energy constraints that are typical of \acp{UAV}.

\subsection{UAVs Integrated {with a} %to the 
Cellular Network}
Differently from the previous scenario, here the \acp{UAV} are assisted by the existing communication infrastructure (e.g., an edge or a cloud) to offload some of the processing or to access complementary knowledge. As in a ``win-win-win" strategy, the collaboration can bring benefits {\em (i)} to the communication and sensing network because the \acp{UAV} act as distributed sensors enhancing context awareness and the overall accuracy, even though offloading might impose additional load to the communication link; {\em (ii)} to \acp{UAV} if the infrastructure supports low-latency communications, and {\em (iii)} to the society at large as it enables immediate \ac{UAV} communication with terrestrial users sharing the same network providing an enhanced connectivity.

In this type of architecture, the \acp{UAV} can be assisted by a remote central engine trained directly on the cloud or on the edge using data collected by the \ac{UAV} sensors. This setup enables the deployment of very simple \acp{UAV} because the computation and the storage are completely delegated to the cloud/edge or to multi-access edge computing servers whose pervasive presence is fostered by the available 5G/6G \acp{BS}. The edge performs customized short-term, low-latency operations, such as local aggregation, storage and caching. Further, the edge can manage some operations which are updated based on \ac{UAV} incoming messages and whose outputs can be transferred to the cloud where all long-term and longer latency operations can be performed.
The \acp{UAV} still perform some simple operations: for example, they collect data through their sensors, update models, compress and encrypt model parameters to be sent out, and take some decisions.
\begin{table*}[!htb]
\centering
\setlist[itemize]{wide, leftmargin=*, noitemsep, after=\vspace*{-\topsep}}
\setlength{\extrarowheight}{3pt}
\setlength{\tabcolsep}{3pt}
\caption{Summary of advantages and disadvantages of different \ac{UAV} network configurations and control methods.}
\label{tab:summary}
\begin{tabularx}{\textwidth}{|p{20mm}|*{2}{>{\compress\RaggedRight\arraybackslash} X |}}
\hline
\rowcolor{aliceblue} \textbf{\ac{UAV} Architectures} & \textbf{Advantages} & \textbf{Disadvantages} \\
\hline
\ac{U2U}
& \begin{itemize}
  \item Fully on-UAV (increased autonomy);
\item Enabler for faster decisions (on-board processing and control);
\item Fault tolerant (higher reliability); 
\item Enabled by ad-hoc \ac{D2D}  or supported by {4G}-\ac{5G} \ac{D2D} communications; %schemes;
\item Flexible and easily reconfigurable;
\item Reduced latency when one-hop is possible.
\end{itemize}
& \begin{itemize}
  \item Limited information available at each UAV (local sensing);
  \item Limited context awareness and estimation performance accuracy;
  \item Increased complexity at the \ac{UAV} side;
  \item Increased energy consumptions;
  \item If multi-hops are present, information can be aged or missed;
\end{itemize} \\
\hline
Integrated with the cellular network
& \begin{itemize}
  \item Enhanced context awareness;
  \item Low latency;
  \item \ac{UAV} can leverage some processing to the edges and be more simple on-board (less complex);
  \item \acp{UAV} are also connected with terrestrial users;
  \item Scalability: appropriate for large swarms;
\end{itemize}
& \begin{itemize}
\item Reduced \ac{UAV} autonomy;
\item Vulnerability to network failures (lower reliability);
  \item Privacy and security issues;
\end{itemize} \\ \hline \hline
\rowcolor{aliceblue} \textbf{\ac{UAV} Intelligence} & \textbf{Advantages} & \textbf{Disadvantages} \\ \hline
Model-based \,\,optimization & \begin{itemize}
  \item Solutions can be derived in a greedy fashion or through closed forms;
\item Enable quick (real--time) decisions;
\end{itemize} & \begin{itemize}
  \item Need accurate models of the environment;
\item Not suitable for heterogeneous and competing tasks;
\item Models may not hold for fast changing environments;
\end{itemize}  \\\hline
Model-free learning & \begin{itemize}
  \item Easy to be reconfigured based on changes of the network or of the mission goals (dynamic environments);
  \item Suitable for heterogeneous and multiple tasks;
%\item Tabular methods are easy to implement and value tables can be stored and updated by an edge;
%\item Deep \ac{RL} helps to accelerate learning by reducing the dimensionality of the problem;
\end{itemize}  & \begin{itemize}
  \item Several episodes might be required (slow convergence);
\item Tabular methods {suffer from} %might occur in 
the curse of dimensionality; % issue;
\end{itemize} \\\hline
%\hline
\end{tabularx}
\end{table*}

%Moreover, even 
If it is %often 
convenient to rely on the existing cellular infrastructure, several concerns should be addressed.
First, one of the strongest points of the \ac{UAV} technology lies in their autonomy in making decisions, for example for navigation, and in their ability to reconfigure quickly the network topology in reaction to temporary or
unexpected events. If the edge takes the role of a pilot for the swarm of \acp{UAV}, the \acp{UAV} become less autonomous.
Secondly, the continuous exchanges with the cloud/edge might introduce substantial data delays, if the cellular network does not support low-latency communications, which in turn leads to their less responsive control \cite{kouhdaragh2020application}. This aspect, together with the vulnerability to network/communication failures, might be extremely limiting when the \acp{UAV} are employed for emergency/surveillance situations.
Privacy is yet another concern that arises when data are shared with the edge to carry out part of the computations.  To avoid privacy issues, federated learning has been recently proposed as an edge-assisted learning paradigm, where only high-level learning parameters are exchanged \cite{yin2020fedloc}.

Table~\ref{tab:summary} presents a summary of the discussion of Sec.~\ref{sec:network}.
\section{UAV Intelligence}
\label{sec:intelligence}

In this section, we provide more insights on \ac{UAV-I}  and, in particular, on methods that can be used for performing mission tasks, where localization is {of primary importance}. %an essential feature. 
As depicted in Fig.~\ref{fig:fig2}, the local \ac{UAV-I} consists of three main blocks: \textit{(i)} an {\em inference engine} that fuses the available information and estimates the state of the environment, e.g., the surroundings and the other \acp{UAV} \cite{lee2020optimization}; \textit{(ii)} a {\em policy \& control module} that allows for learning and updating a policy, i.e., a rule, for taking decisions and actions; \textit{(iii)} a short-term memory that is an internal repository of knowledge, immediately available for \ac{UAV} urgent computations, especially useful in full autonomous settings. Such modules perform disjoint operations, but they are strictly intertwined according to the level of \ac{UAV} autonomy. %In the 
Next, we briefly describe the modules. % by module.

\subsection{Inference Engine}
\label{sec:inference}

The inference problem consists of estimating the state of the phenomenon of interest from observations collected by sensors. 

When the state parameters are described by a statistical model, the estimation can be performed within the Bayesian framework. In many practical cases, the state-observation pair can be represented by a probabilistic state-space model that includes \cite{Gus:J10,DunEtAl_J20}: \textit{(i)} a {\em prior model} that quantifies the initial knowledge about the state; \textit{(ii)} a {\em measurement model} that relates the state to the measurements by the likelihood of the state; \textit{(iii)} a {\em transition model} that represents the temporal evolution of the state.
Given this state-space model, Bayesian filtering produces the predictive \ac{pdf} of the state based on {past collected measurements} and, then, updates it to the filtering \ac{pdf} of the state using new measurements and by applying the Bayes' rule.

Kalman and particle filters are common methods for solving sequential estimation problems and they can also be used in a distributed manner. Probabilistic mapping is also useful to reconstruct a binary map of the environments \cite{thrun2002probabilistic}.  In cooperative scenarios, all these methods
can be used in a distributed manner making use of factor
graphs, belief propagation or distributed \ac{PF} \cite{dardari2015indoor,win2011network}.
%In the following, we review the principles behind policy estimation.

\subsection{Policy \& Control Engine}
\label{sec:policy}

The interaction between a \ac{UAV} and the environment, shown in Fig.~\ref{fig:fig2}, can be described under the umbrella of Markov decision processes (MDPs) %\ac{MDP} 
or as partially observable %\ac{MDP} 
if the state is unknown. 
An MDP % \ac{MDP} 
is defined by a tuple containing the space of possible states/observations, actions and rewards, and the probability of state transition. 

Notably, \acp{UAV} act as local agents, and the state represents the available knowledge about the environment. Each \ac{UAV}, then, interacts with the environment through the chosen actions and, in turn, it receives rewards defined by a metric that measures the goodness and the proficiency of a taken action. In this sense, a policy is a function that maps the observations to the actions, and, hence, {\em learning} is the process that allows inferring such policy.

In the following, we discriminate between methods that rely on accurate models of the environment, solved using classical optimization techniques, from methods that learn from experience using artificial intelligence/machine learning (AI/ML) tools. Table~\ref{tab:summary} provides  %reports 
a summary of the comparison between model-based and model-free methods for policy evaluation.
\medskip

\textbf{Model-based methods}: Classically, this problem is solved by using model-based optimization, e.g., nonlinear programming, where a single cost function or a limited set of them is minimized either in a greedy fashion or by closed-form solutions, allowing for taking quick decisions. 

For example, in \cite{guerra2020access, zhang2020self} the \ac{UAV} navigation problem is described as the minimization of the posterior covariance matrix on target estimation subject to \ac{UAV} anti-collision and mobility constraints. In this case, the cost function to be minimized can be analytically derived because observation and transition models are both {known and} Gaussians. 
Then, to solve the problem, a projected steepest gradient descent algorithm is used whose major steps are \cite{zhang2020self}: \textit{(i)} analytical derivation of the gradient of the cost function with respect to the \ac{UAV} positions; \textit{(ii)} computation of an initial solution with the steepest gradient descent of the cost function; \textit{(iii)} derivation of the projection matrix to constrain the initial solution onto the tangent space of the potentially violated constraints; \textit{(iv)} corrections to compensate for the nonlinearity effect of the constraints ({\em nonlinearity correction}); \textit{(v)} derivation of the projected control for \acp{UAV}. 

Unfortunately, when the dynamics of the environment is only partially known, model-based approaches might not be sufficiently accurate. In this context, \ac{ML} techniques, {i.e., model-free methods,}  become essential for solving a large number of heterogeneous and, sometimes, competing tasks (e.g., mission duration, storage capacity, communication requirements) \cite{kouhdaragh2020application}.  However, when empirical methods are not available, \ac{ML} strategies offer the possibility of inferring a policy in a data-driven way by interacting with the environment through on-board sensors and message (data) exchanges. 
\medskip

\textbf{Model-free methods}: {Typical \ac{ML} approaches are usually distinguished in} {\em supervised learning}, that learns from a training set of labeled examples, and {\em unsupervised learning}, that extracts hidden features from unlabeled datasets. However, they do not usually consider the 
whole problem of an application-driven agent that interacts with an uncertain environment to maximize a reward. %{\color{red} DD: Perhaps I'm wrong, but can RL be considered as a subset of ML, or not?  }
% unlike \ac{RL} \cite{lee2020optimization}. 
{Instead, in}  \ac{RL}, {a third \ac{ML} paradigm}, the learning is carried out through interactions (``{\em trial-and-error}") and actions are selected to maximize the sum of the discounted rewards over the future (``{\em delayed rewards}") \cite{lee2020optimization,ahmed2021reinforcement}. 

The policy can be estimated through a {\em direct} or {\em indirect learning} approach. In direct learning, the policy is represented by a neural network with observations as inputs and actions as outputs, so that the policy is directly inferred by approximations ({\em policy search}). Thus, learning simply consists of adjusting the parameters of the neural network to find the optimal input/output relationship. Possible algorithms to perform such updates are represented by policy gradient methods \cite{lee2020optimization}. Unfortunately, such methods might have slow convergence and they can be stuck on local maxima.

Differently, indirect learning is a mapping of a state-action pair into a value function, defined as the expected return, i.e., a discounted reward sequence over a certain time horizon. Since the policy should return an action to the \ac{UAV}, it assesses the values 
for every possible action and decides either for the optimal action ({\em exploitation}) or a random one ({\em exploration}).

Implementations of this policy include approaches based on dynamic programming, Monte Carlo methods, and temporal-difference learning \cite{lee2020optimization}.
As an example of off-policy \ac{TD} control algorithm, $Q$-learning is often used in grid-world environments with a finite set of states and  possible actions \cite{lee2020optimization}. The policy is learnt run-time while the \ac{UAV} is navigating the environment.  \ac{TD} methods  use a generalized policy iteration %\ac{GPI} 
mechanism to alternatively estimate the optimal policy and the optimal $Q$-value. 

The advantages of using \ac{TD} methods instead of Monte Carlo or dynamic programming is that there is no need of a model for the environment's dynamics and an update of the return is made at each time step. For discrete states and actions, the $Q$-value can be represented by a $Q$-table that, at each time instant, is updated based on the collected rewards.
Obviously, the tabular approach is practical when the number of states and actions is discrete and small, while it becomes not feasible for continuous state and action spaces (due to the \textit{curse of dimensionality}).
In high-dimensional problems, deep \ac{RL}  permits to reduce their dimensionality (e.g., complexity) by exploiting the adoption of deep neural networks
\cite{lee2020optimization}.

Table~\ref{tab:summary} reports a summary of the discussion of Sec.~\ref{sec:intelligence}.

\section{Case Studies}
\label{sec:example}
 We now propose two examples of localization-based applications in order to provide a discussion about some of the aspects previously discussed, % proposed, 
 such as offloading the \ac{UAV-I}, and the operation under the possible availability of an accurate model of the environment. 
 
\begin{figure*}[t!]
 \centering
 \includegraphics[width=0.9\linewidth,draft=false]
 {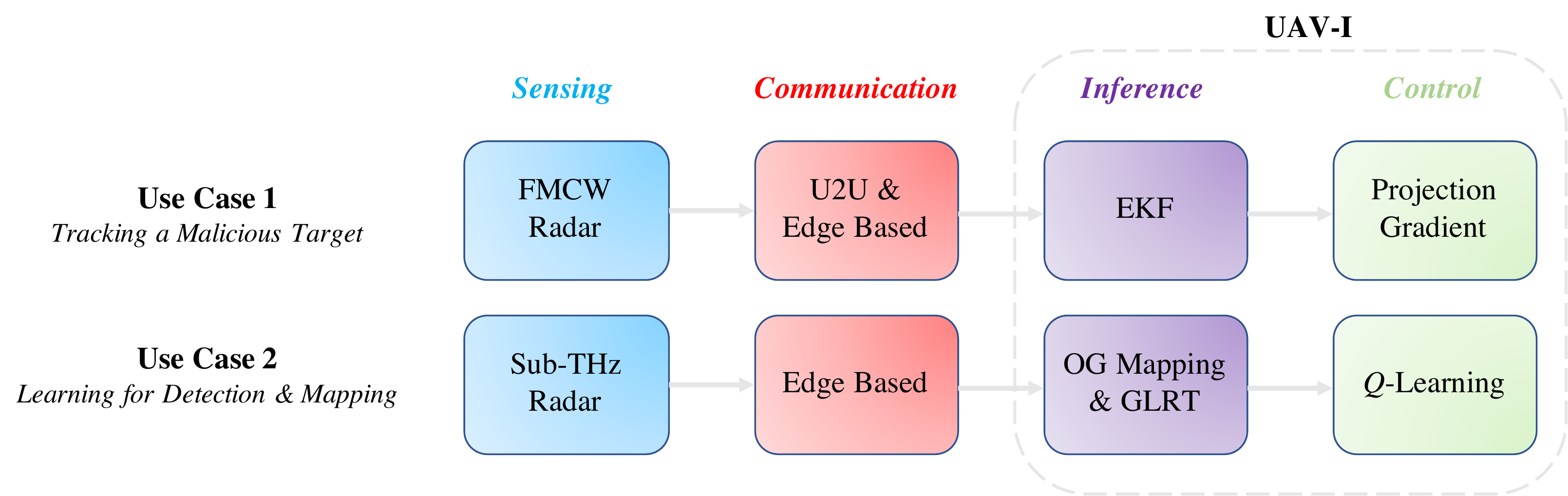}
 \caption{Considered use cases and related tools.}
 \label{fig:diagram}
 \end{figure*}

 {\emph{Use Case $1$}} deals with the problem of tracking a malicious moving target by a swarm of \acp{UAV}, whereas, {\emph{Use Case $2$}} studies the exploration of an unknown environment together with the detection of a target that can be either collaborative or unintended, e.g., revealed by spectrum-sensing cognitive radio techniques. Note that, in both applications, {the mission has to be completed by a given deadline.}
 %time is a limited resource and a limitation to successfully complete the mission.
 Fig.~\ref{fig:diagram} depicts the implementations of the main modules at each \ac{UAV} for the two use cases.
 
\begin{table}[t!]
\caption{Complexity for Use Case 1.} 
\label{tab:EKFNAVcost}
\makebox[\linewidth]{
\begin{tabular}{l|l|l|l}
{\textbf{Operations}} &  \multicolumn{3}{c}{{\textbf{Parameters}}}  \\ \hline
 &  {\textbf{State}} & {\textbf{Number of}} & {\textbf{Number of}}  \\ 
& {\textbf{Dimension}} & {\textbf{Measurements}} & {\textbf{Constraints}}  \\\hline
\hline 
 \multicolumn{4}{c}{ \textit{State Inference (EKF)}} 
\\ 
\hline\hline
\rowcolor{lightpastelpurple}  Predicted state mean & Quadratic & - & -\\ \hline
\rowcolor{lightpastelpurple}  Predicted cov. matrix&  Cubic & - & -  \\ \hline
\rowcolor{lightpastelpurple} Kalman Gain & Linear &  Cubic & - \\ \hline
\rowcolor{lightpastelpurple} State mean & Linear & Linear & - \\ \hline
\rowcolor{lightpastelpurple} State covariance  & Quadratic & Quadratic & - \\\hline
\hline
  \multicolumn{4}{c}{ \textit{UAV Control  (Gradient-based)}}    \\\hline
 \hline 
\rowcolor{lightgreen}  Nonlinearity correction & Linear &- & Cubic  \\ \hline
\rowcolor{lightgreen}  Projection matrix & Quadratic & -  & Linear \\ \hline
\rowcolor{lightgreen} Projected control & Quadratic & - & -  \\ \hline
% Control signal & $0$\\ \hline
\end{tabular}
}
\end{table}

\subsection{Use Case 1: Tracking a Malicious Target}

According to Fig.~\ref{fig:diagram}, each \ac{UAV} implements the following operations for target tracking and navigation:
\begin{itemize}
\item {\em Sensing}: Among other sensors, each \ac{UAV} is {equipped} %embedded
with a low-cost millimeter-wave \ac{FMCW} radar collecting radar echoes in the form of range measurements. This operation cannot be offloaded to an external entity.
\item {\em Communication}: Each \ac{UAV} can decide whether to communicate with other neighboring \acp{UAV} of the network or to be assisted by an edge. The payload data consist of the \ac{UAV} ID, a timestamp, a range measurement, and the \ac{UAV} position.
\item {\em UAV-I}: This module consists of:
\begin{itemize}
\item	{\em Inference Module}: the \ac{UAV-I} runs an \ac{EKF} for inferring the state;
\item {\em Control Module}: the \ac{UAV} adopts a low-complexity navigation algorithm for minimizing the target localization error accounting for anti-collision constraints. 
In this sense, we rely on a model-based method solved using a projection steepest gradient descent algorithm.
\end{itemize}
\end{itemize}

\subsubsection{Computational Complexity Analysis}%Computational Complexity Analysis}
\label{sec:complexity}
{We aim here at %practically 
understanding the complexity {for %entailed to 
carrying out} the various operations required to complete the mission for which the \acp{UAV} are responsible. %In turns, 
This analysis is essential for deciding how to distribute the \ac{UAV-I} when a network for localization-based applications is considered.} Because the first two operations cannot be delegated off-board, we only focus on the computational analysis of the {\em inference} and {\em policy} learning modules. The considered \ac{EKF} algorithm is composed of two major steps, the {\em prediction}, with a cubic dependence on the number of states, and the {\em update}, which has a cubic dependence on the number of measurements due to the Kalman gain derivation, and a quadratic dependence of both states and measurements in the state covariance matrix derivation (see Table~\ref{tab:EKFNAVcost}). 
The projection gradient algorithm for navigation is derived in closed form to reduce the complexity to only few multiplications that depend on the number of violated constraints \cite{guerra2020access}.
%For this analysis, we consider an \ac{EKF} algorithm that is composed of two major steps, the prediction, with a cubic dependence on the number of states, and the update, which has a cubic dependence on the number of measurements due to the Kalman gain derivation, and a quadratic dependence of both states and measurements in state covariance matrix derivation (see Table~\ref{tab:EKFNAVcost}). The projection gradient algorithm for navigation is derived in closed form to reduce the complexity to only few multiplications that depend on the number of violated constraints \cite{guerra2020access}.

%
\begin{figure}[t!]
 \centering
 \includegraphics[width=0.95\linewidth,draft=false]
 {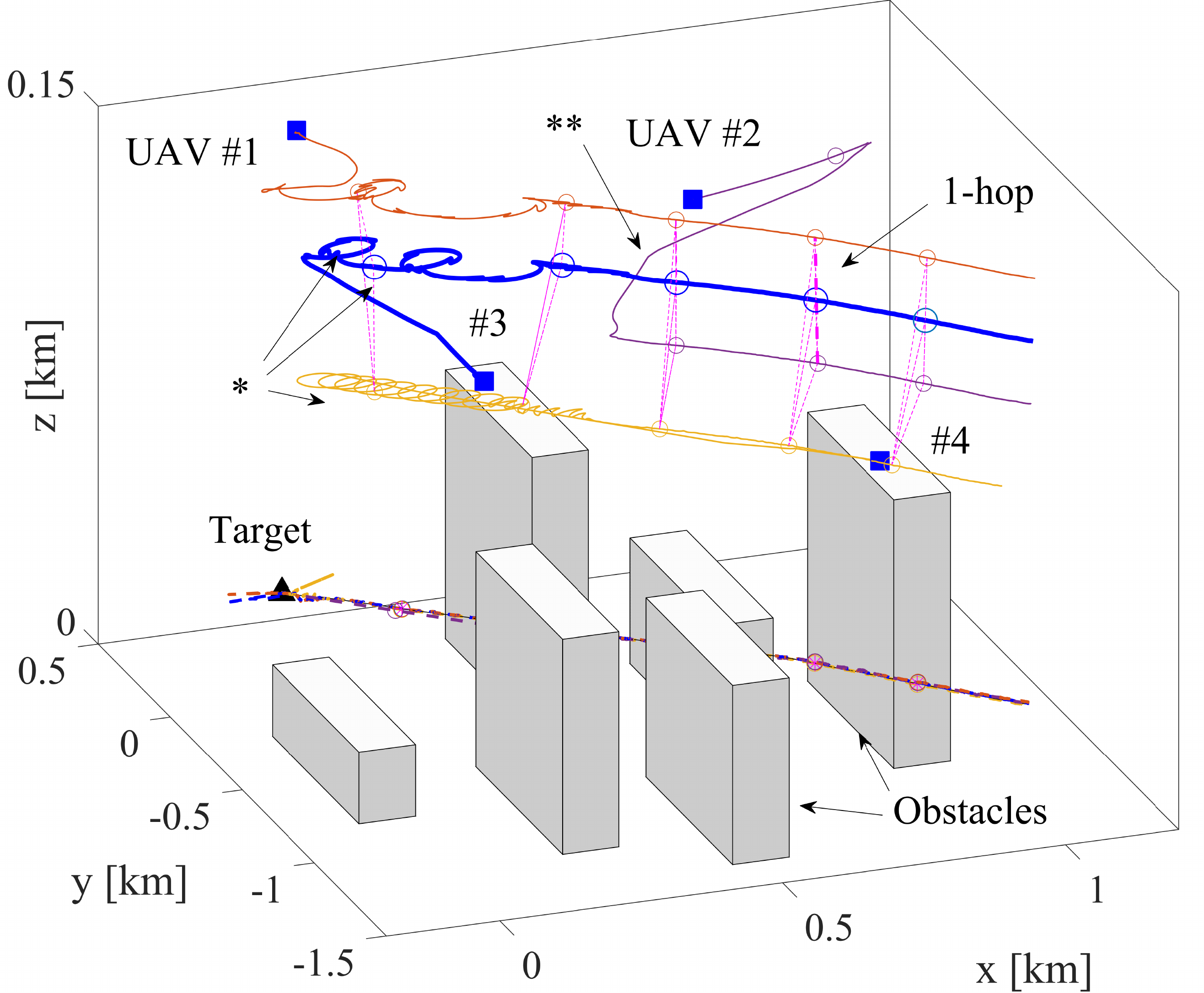}
 \caption{Localization scenario for \ac{U2U} communication.}
 \label{fig:fig3}
 \end{figure}
\begin{figure}[t!]
 \centering
 \includegraphics[width=0.95\linewidth,draft=false]
 {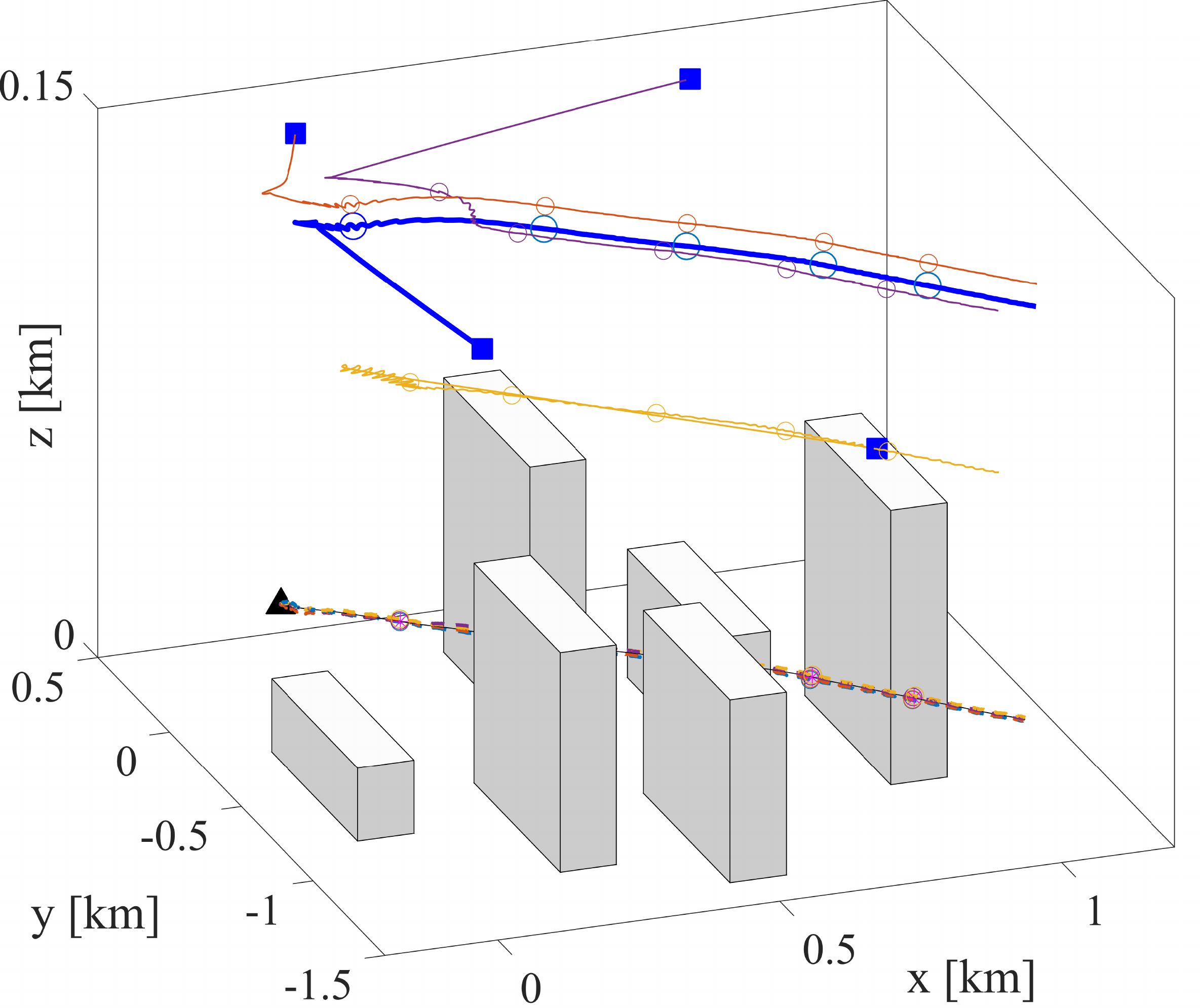}
 \caption{Localization scenario for edge-based communication.}
 \label{fig:fig4}
 \end{figure}
\begin{figure}[t!]
 \centering
 \includegraphics[width=0.94\linewidth,draft=false]
 {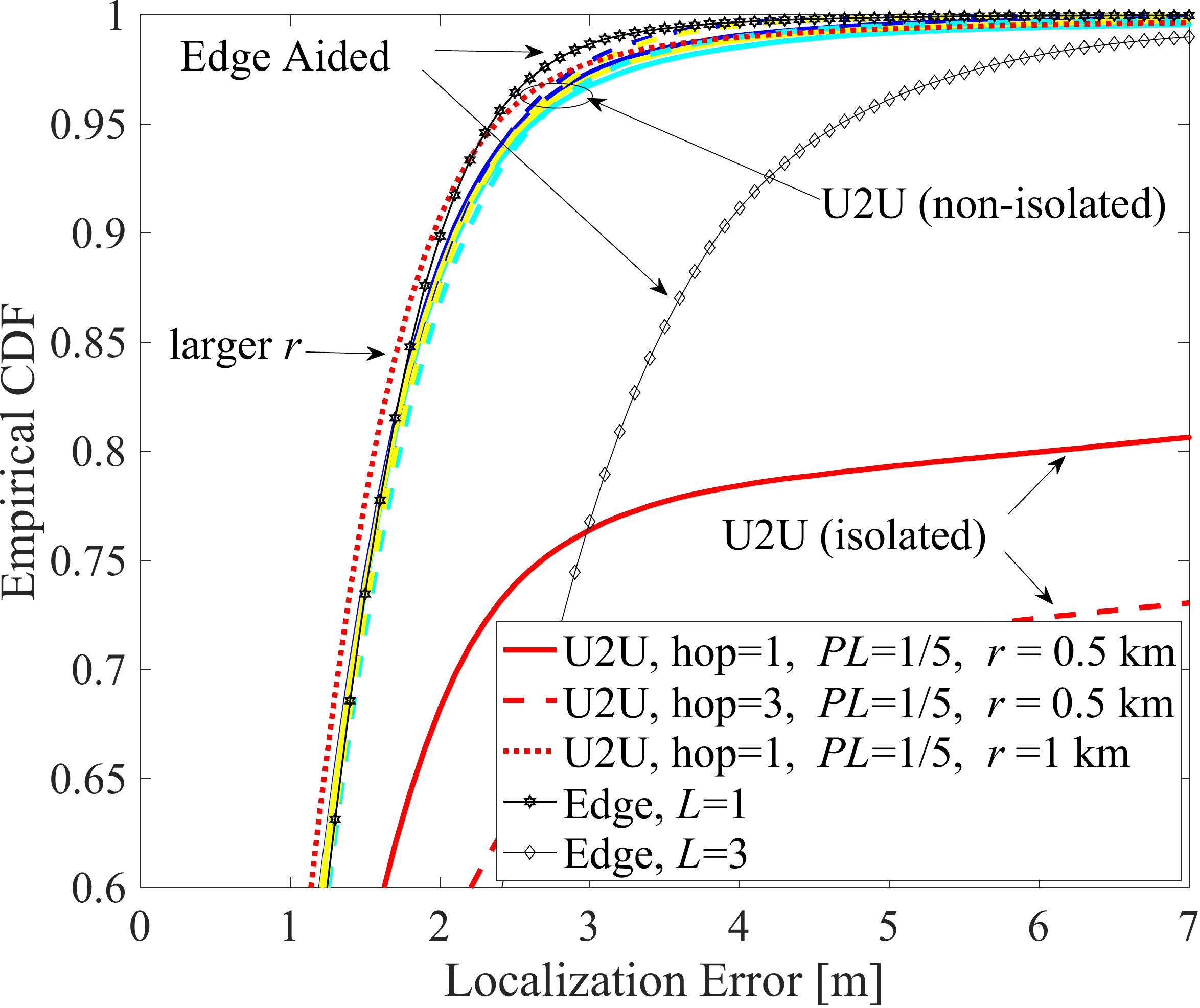}
 \caption{Empirical {CDF} as a function of the localization error. $L$, $PL$ and $r$ indicate the latency (delay in packet reception) in terms of time steps, the packet loss and the communication range, respectively. For \ac{U2U}, different colors indicate a different \ac{UAV}, i.e., the red one stands for \ac{UAV} \#2, whereas the line style meaning is indicated in the legend.}
 \label{fig:fig5}
 \end{figure}

\subsubsection{ Performance vs. Complexity Offloading}
\label{sec:performance}

We now consider a scenario as in Figs.~\ref{fig:fig3}-\ref{fig:fig4}, where a swarm with four \acp{UAV}, with initial locations in the blue squares, tracks a target initially located in the black triangle. The actual target trajectory is displayed with a black line and the estimates of the \acp{UAV} with colored dashed lines. The same colors are also used to represent the trajectories of the \acp{UAV}. 
%In this case, we consider each \ac{UAV} equipped with a \ac{FMCW} for tracking.

Two different network architectures were investigated. The first was %considers 
a network with fully autonomous \acp{UAV}, communicating through a multi-hop \ac{U2U} scheme with a maximum communication range of $r=0.5\,$km or $r=1\,$km, and up to three hops. In this multi-hop scheme, a \ac{PL} of $1/5$ was assumed. 
%Instead, 
The second scheme was edge-assisted: all the sensed data were sent to the edge and, for our setting, delayed by $L$-time steps {(accounting for edge access/processing delay)} and a negligible \ac{PL} thanks to the high cellular network reliability.
The \ac{UAV} speed was varying by $\left[8, 10\right]\,$m/step, while the target was moving more slowly along a straight line with random accelerations and with an initial speed of $1\,$m/step. This critical application required the \acp{UAV} to be capable of tracking the ``malicious" target position in the shortest possible time. %, only $10$ time instants are considered.
{Note that the two adopted communication schemes affected the \ac{UAV} trajectory. In fact,} Fig.~\ref{fig:fig3} shows that, in the \ac{U2U} configurations, the \acp{UAV} followed circled trajectories, indicated with ``*", when the number of available measurements at each \ac{UAV} was not sufficient to make a proper decision on navigation, and they had to rely only on their own sensed data. Moreover, \ac{UAV} \#$2$ was isolated during the first instants of the mission. On the contrary, when the network was fully connected with a one-hop, indicated with two stars, the \acp{UAV} flew in a team formation. The same was valid when the communication was assisted by the edge, shown in Fig.~\ref{fig:fig4}, where the \acp{UAV} were never isolated, and they approached the target more quickly.

Figure~\ref{fig:fig5} provides the empirical \ac{CDF} defined as the number of times the absolute positioning error is lower than a desired threshold, averaged over Monte Carlo iterations and time. The localization error of the edge-assisted network was below $2\,$m in almost $90$\% of cases if the cellular infrastructure guaranteed low latency (i.e., $L=1$). Conversely, with one-hop \ac{U2U} communication the same accuracy could be achieved if the communication range was large enough (e.g., $r=1$ km). \ac{U2U} schemes could be more accurate than the edge-assisted schemes if the latter introduced some delay $L>1$ because they processed fresh data with very low latency.
{However, \ac{U2U} schemes might suffer from information losses in case some \ac{UAV} gets isolated because of large distances.}
%they can suffer from reduced information because \acp{UAV} can be isolated if they are too distant from each other and some packets are lost. 
Edge-based solutions, on the other hand, can rely on more data and, thus, more information and better coverage, and they can be based on more complex computations. The price is a slower convergence to the desired localization accuracy, if the edge has high latency. % and privacy issues.

\subsection{Use Case 2: Real-time Learning for Environment Mapping \& Detection of a Collaborative Target}

We now provide a use case for model-free \ac{UAV} navigation where a swarm of \acp{UAV} navigated an unknown indoor environment with the intent to detect a target while reconstructing a map of the surrounding ambient.  %\ac{RL} is used to solve a joint navigation, detection and environment mapping problem. 
%In fact, differently from the previous example where a model was available to solve the problem, we consider a real-time learning problem, where \acp{UAV} navigate an unknown indoor environment, and they have to detect a target while reconstructing a map of it. 
{Such a scenario can represent a search--and--rescue situation of a cooperative target, or the detection of a malicious user that hides behind an obstacle and whose communication with other users is revealed through ad-hoc cognitive radio techniques.
}
Note also that this analysis %here reported
extends the work in \cite{GueGuiDarDju:C21}, which is valid for only one \ac{UAV}.

\begin{table}[t!]
\caption{Complexity for {\em Use Case 2}.} 
\centering
\scalebox{0.9}{
\label{tab:qlearn}
\makebox[\linewidth]{
\begin{tabular}{l|l|l|l}
{\textbf{Operations}} &  \multicolumn{3}{c}{{\textbf{Parameters}}}  \\ \hline
 &  {\textbf{State}} & {\textbf{Number of}} & {\textbf{Number of}}  \\ 
& {\textbf{Dimension}} & {\textbf{Measurements}} & {\textbf{Actions}}  \\\hline
\hline 
 \multicolumn{4}{c}{ \textit{State Inference (Occupancy Grid)}} 
\\ 
\hline\hline
\rowcolor{lightpastelpurple}  Likelihood computation  & - & Linear & -\\ \hline
\rowcolor{lightpastelpurple}  Log-odd update & Linear & - & - 
 \\\hline
\hline
  \multicolumn{4}{c}{ \textit{UAV Control ($Q$-Learning)} for each episode}  \\\hline
 \hline 
\rowcolor{lightgreen} Generate a random $\epsilon$ & - &- & -  \\ \hline
\rowcolor{lightgreen} Evaluation of rewards  & Linear &- & -  \\ 
\rowcolor{lightgreen}
\hline
 Find the maximum  (if greedy)  & - &- & Linear \\ 
 \rowcolor{lightgreen}\hline
  Update $Q$-table  & Linear &- & Linear \\ 
\rowcolor{lightgreen}\hline
\end{tabular}
}
}
\end{table}
%

%In this case, 
The main modules of Fig.~\ref{fig:diagram} at each \ac{UAV} were:
\begin{itemize}
\item {\em Sensing}: For mapping purposes, a sub-\ac{THz} multi-antenna radar collecting a range-angle (energy) observation matrix was embedded on-board the \acp{UAV}.
\item {\em Communication}: %In this case,
The communication only took place with an edge for sharing some learning parameters and offloading some processing related to decision--making for \ac{UAV} navigation.
\item {\em UAV-I}: For this module, we had %have
\begin{itemize}
\item	{\em Inference}: In the considered example, the \ac{UAV-I} ran two algorithms: \textit{(i)} an \ac{OG} for mapping and a \ac{GLRT} for target detection \cite{GueGuiDarDju:C21}; 
\item {\em Control}: The policy for navigation was learnt run-time with $Q$-learning, a model-free tabular method based on the construction of a $Q$-table \cite{lee2020optimization,GueGuiDarDju:C21}. 
\end{itemize}
%Because the available models are enough accurate, we rely on a model-based method solved using a projection steepest gradient descent algorithm, as described in Sec.~\ref{sec:policy}.
\end{itemize}

\subsubsection{Computational Complexity Analysis}
\label{sec:UC2computAnalysis}
Table~\ref{tab:qlearn} summarizes the complexity of the {\em inference} and {\em control} modules of {\em Use Case 2}. We assumed that the \ac{GLRT} is less complex with respect to the implemented occupancy grid in terms of operations (e.g., it can be implemented with a look-up table according to the \ac{SNR}), and thus we report the complexity in terms of number of measurements and states (i.e., of the number of cells in which the environment is discretized). Further,  we assume that the map belief does not account for inter-cell correlations.  Finally, the $Q$-learning also depends on the number of episodes needed to achieve the final state \cite{koenig1993complexity}. 
Although the main operations have a linear dependence on the state dimension, the number of cells can be huge, even greater than $10,000$ states for large indoor environments. Hence, with low-complexity \acp{UAV}, it would not be feasible to perform state estimation or the $Q$-table update exclusively on-board.
 
\subsubsection{Performance vs. Learning Time}
%{\color{red} DD: I would move all the description of the system at the beginning of section B. After we have described the system, we analyze the complexity and the performance vs learning time trade off. } 
%We now consider a scenario with two \acp{UAV} for accomplishing the task to detect a collaborative user while reconstructing a copy of its surrounding environment (namely, environment mapping). Such a case study extends the one reported in \cite{GueGuiDarDju:C21} for a single \ac{UAV}.

\begin{figure}[t!]
 \centerline{
 \includegraphics[width=0.7\linewidth,draft=false]
 {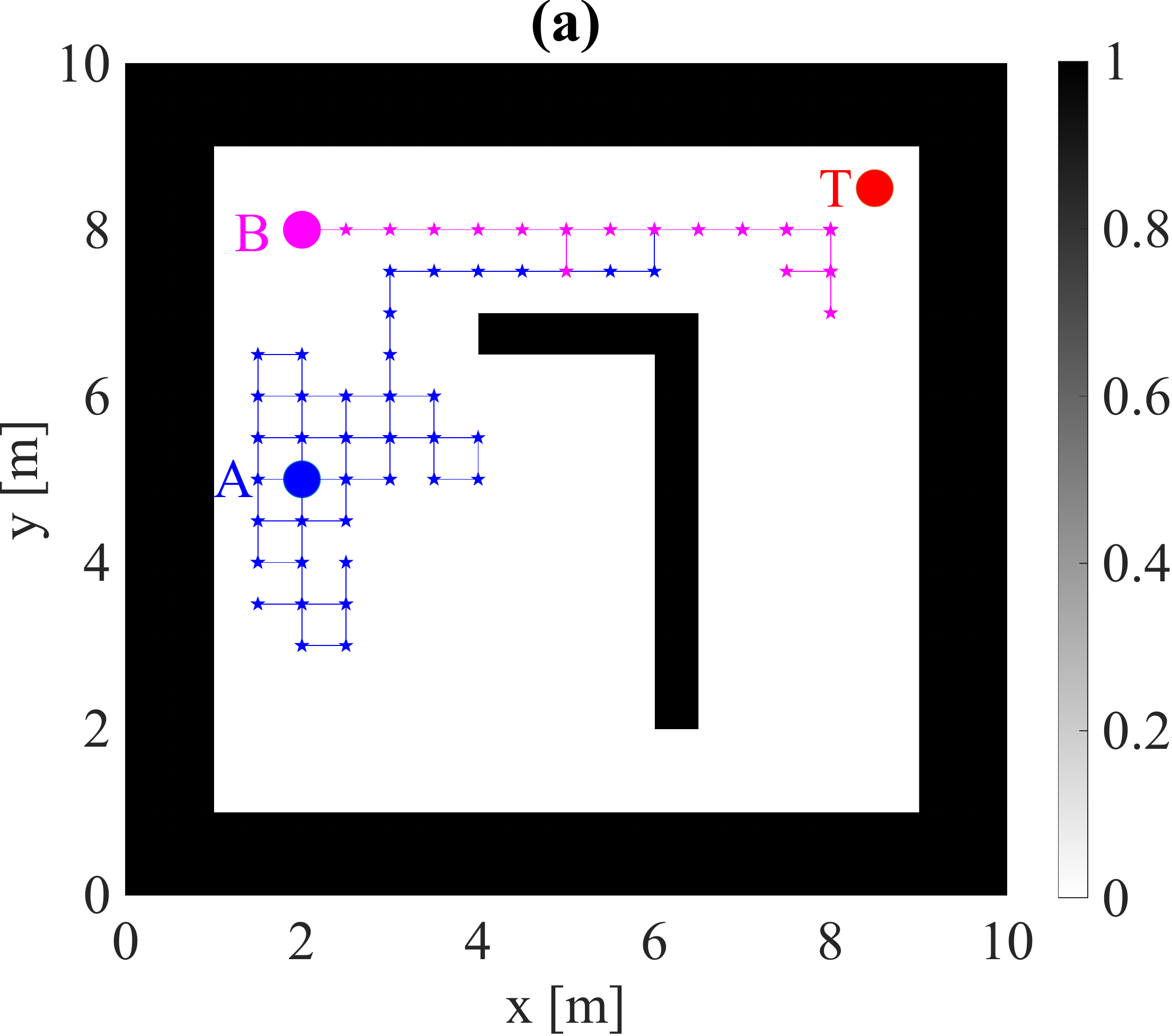}}
  \vspace{0.4cm}
  \centerline{
 \includegraphics[width=0.5\linewidth,draft=false]
 {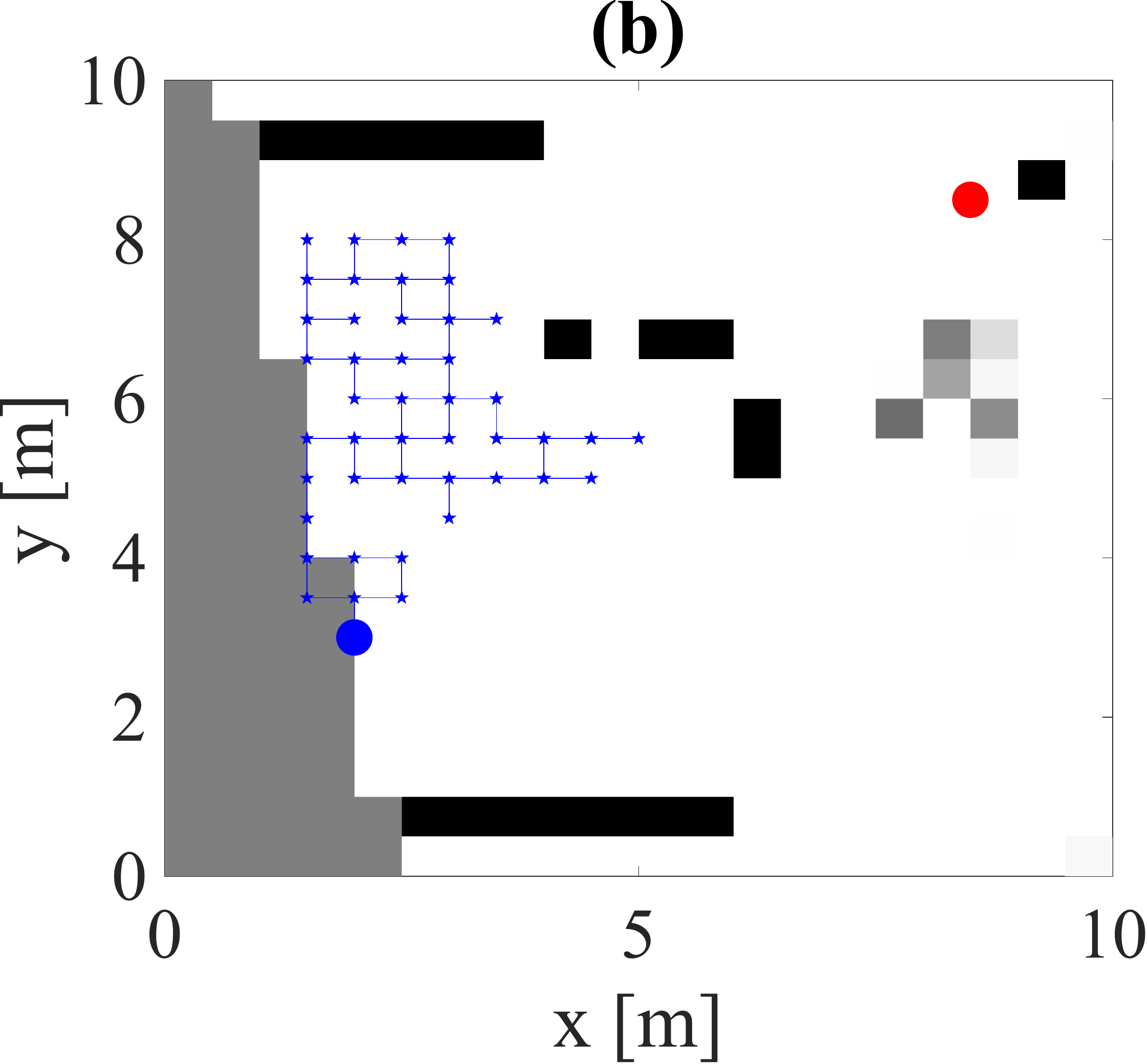}\includegraphics[width=0.5\linewidth,draft=false]
 {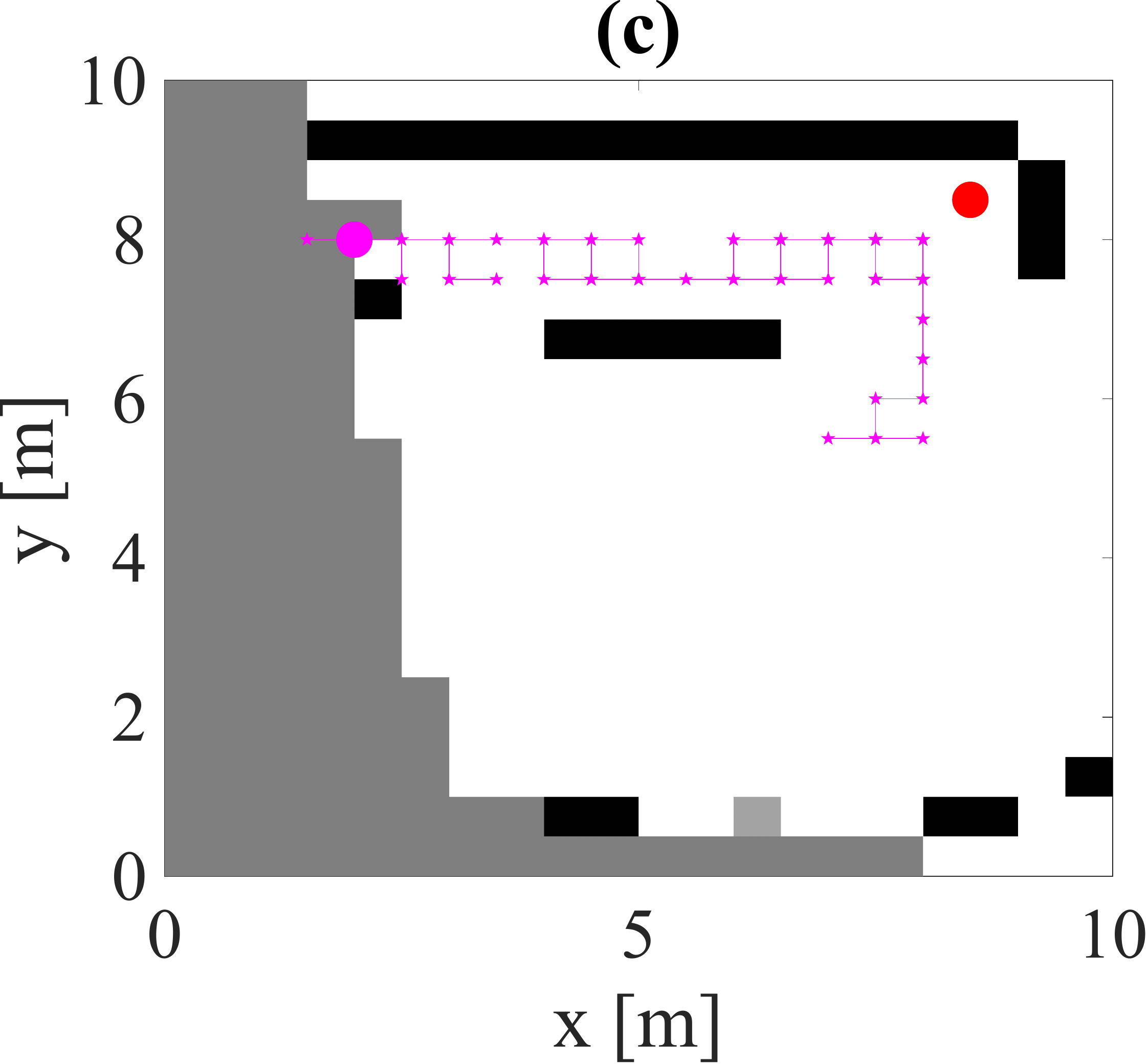}}
 \vspace{0.4cm}
   \centerline{
 \includegraphics[width=0.5\linewidth,draft=false]
 {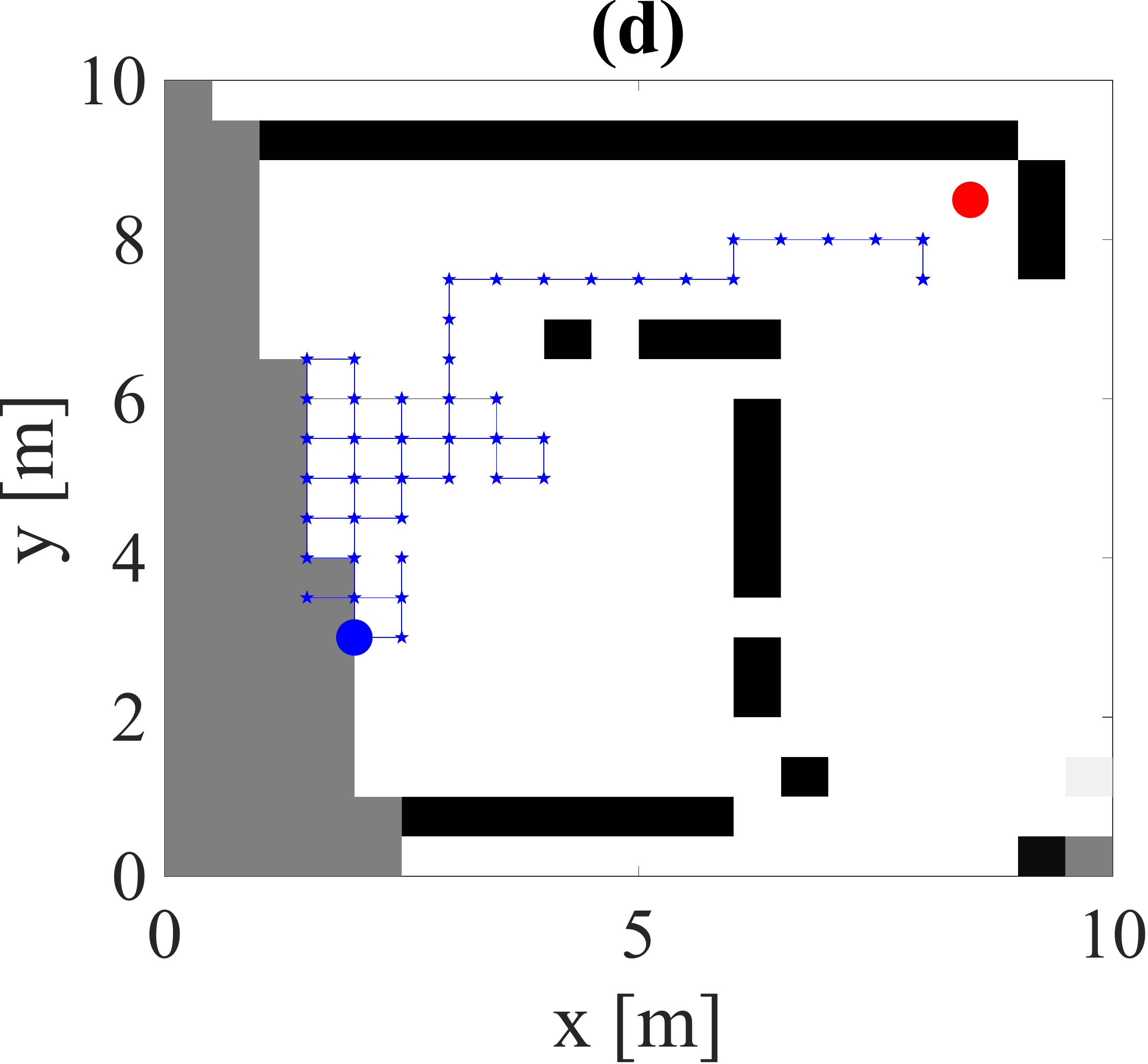}\includegraphics[width=0.5\linewidth,draft=false]
 {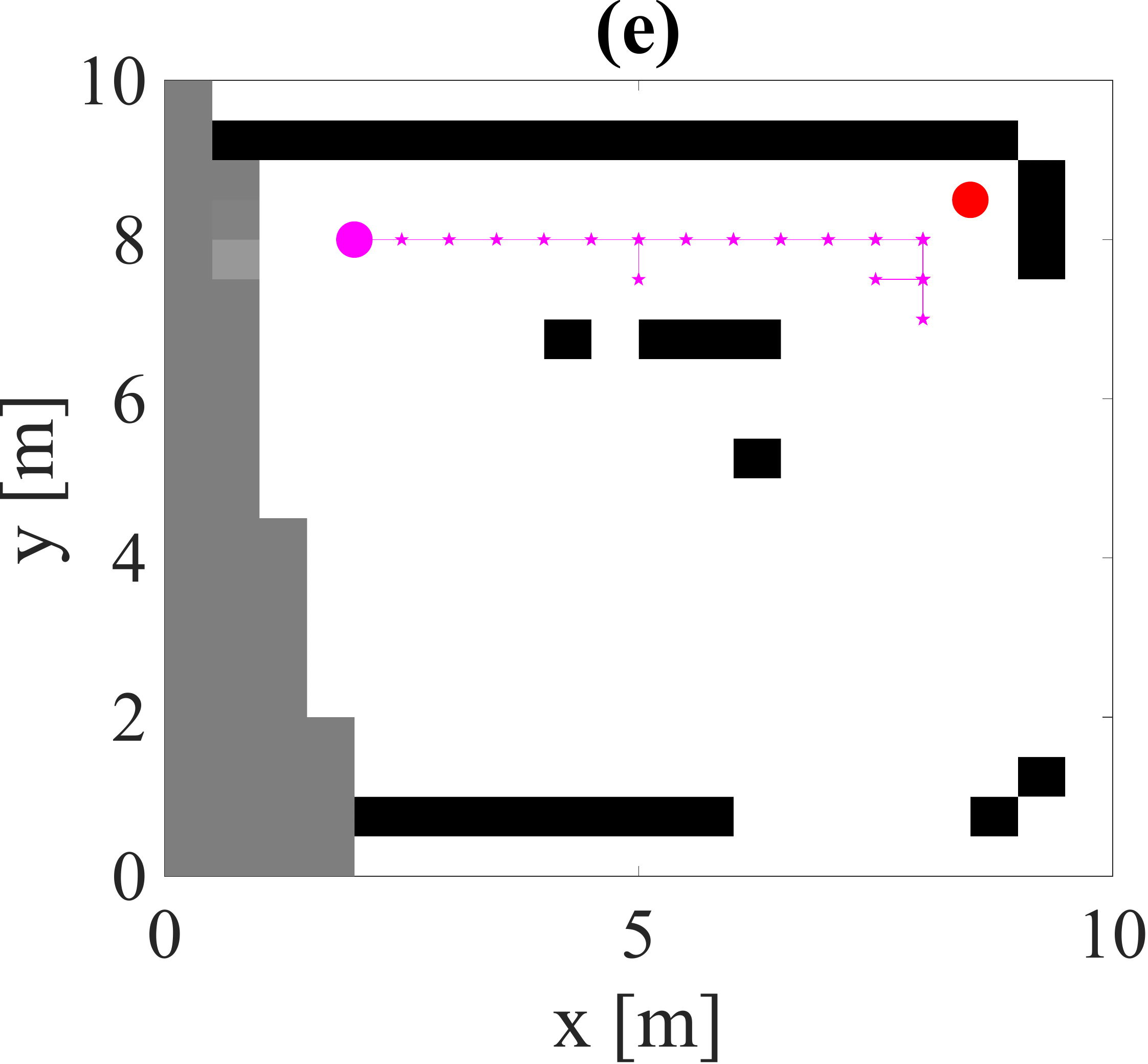}}
 \caption{Examples of estimated trajectories and maps for the first (b,d) and second agent (c,e) and for the first (b,c) and last (d,e) episodes. The reference map is in sub-figure (a). Blue, magenta, and red markers indicate the initial position of the first and second UAV, and the target position, respectively. }
 \label{fig:fig7}
 \end{figure}

We now describe the mapping and detection performance of a network of two \acp{UAV} moving in an environment depicted in Fig.~\ref{fig:fig7}-(a).
The \acp{UAV} were initially assumed to be in positions  $(2,\,3)\,$m and  $(2,\,8)\,$m (blue and magenta markers of Fig.~\ref{fig:fig7}-(a)), respectively, and they moved with steps of $0.5\,$m, equal to the cell width. The task of the network was to detect the target located on the upper east side of the map, in  $(8.5,8.5)\,$m,  and depicted with a red marker. Mapping of the environment is functional to the realization of this task.

Since each \ac{UAV} was equipped with a sub--\ac{THz} radar working at $140\,$GHz, the observation model accounted for the \ac{THz} scattering model of \cite{ju2019scattering, GueGuiDarDju:C21}. In particular,  we set a scattering coefficient to $0.5$ (rough surface), the length of the scattering object to $0.5$, and the width of the scattering lobe to $1$ \cite{ju2019scattering}.
%Other parameters 
We considered an \ac{EIRP} of $30\,$dBm, a receiver noise figure of $4\,$dB, and $1\,$GHz signal bandwidth. The mapping was performed by a radar equipped with an array of $100$ antennas with $10$ steering directions for a (semi--plane) scan of the environment.  For other mapping parameters, we refer to \cite{GueGuiDarDju:C21}. 
\begin{figure}[t!]
 \centering
 \includegraphics[width=0.95\linewidth,draft=false]
 {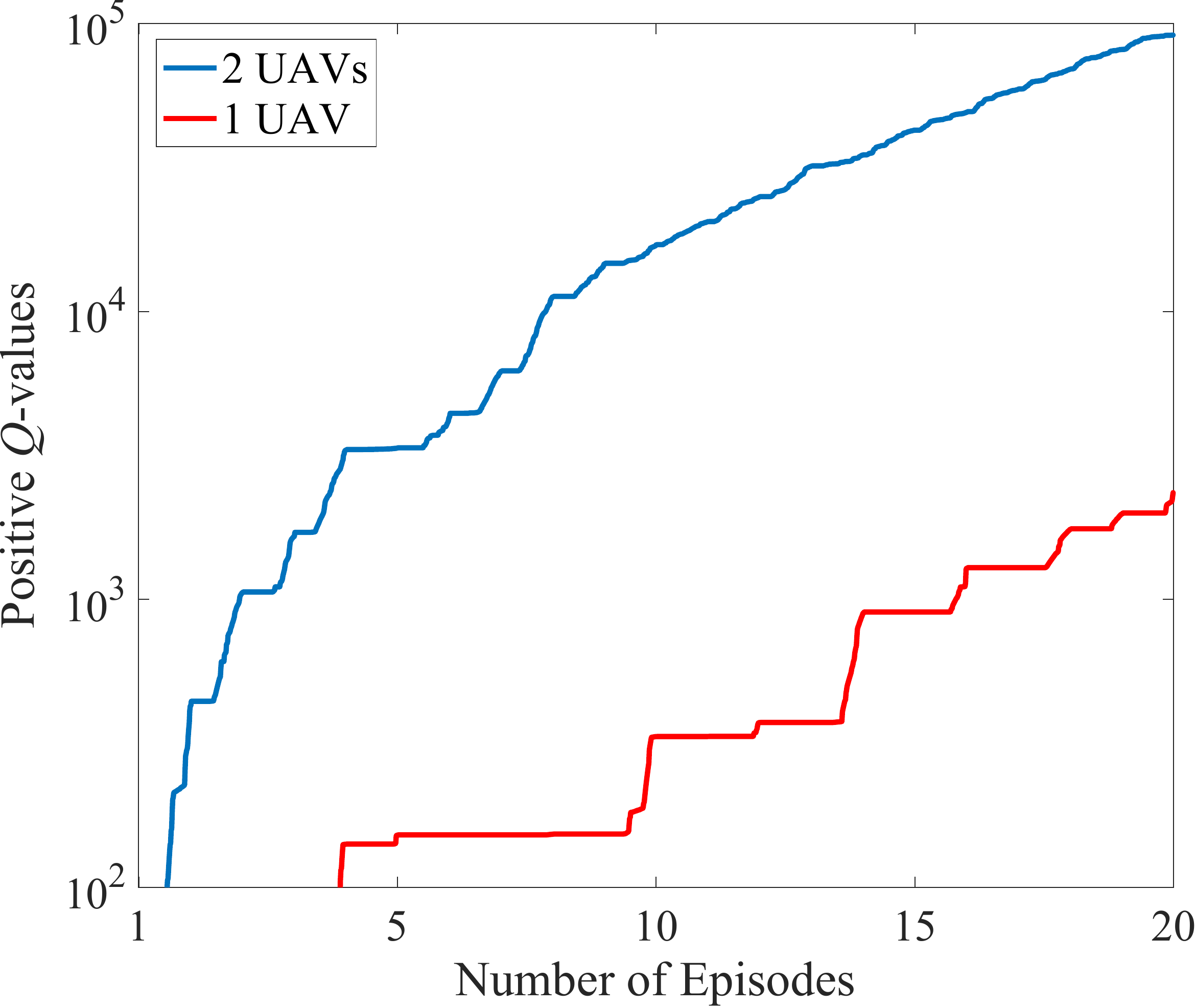}
 \caption{Positive $Q$-values for the detection and mapping scenario of Fig.~\ref{fig:fig7} with a network of one or two \acp{UAV}.}
 \label{fig:fig8}
 \end{figure}

For the detection task, we considered an antenna with $0\,$dBi gain, a reading range of $7.35\,$m and a target always present in the environment. 
A \ac{GLRT} was used to decide whether the target is present or not with a threshold accounting for a desired false alarm probability of $10^{-3}$ \cite{GueGuiDarDju:C21}.  
We fixed  the mission time and the number of episodes to $200$ and $20$, whereas the discount and learning rates to $0.99$ and $0.9$, respectively. We adopted a $\epsilon$-decay strategy where the probability of taking a random action decreased with time.

In regards to inter--UAV coordination, the problem of multi-agent \ac{RL} has already been deeply investigated in \cite{lee2020optimization}. Following our analysis on computational complexity, it was evident that the control, i.e., the store and the update of the $Q$-table, should be delegated to the edge. In fact, given the high-dimensional representation of the environment and the presence of multiple \acp{UAV} with different views of it, the edge served as a leader in fusing the received information, in storing and updating a global $Q$-table. Thanks to this centralized approach, both agents had better ambient awareness and they could infer a global policy  according to $Q$-learning.

Figure~\ref{fig:fig7} shows the \ac{UAV} trajectories with blue and magenta markers for the two \acp{UAV}, and for the first and last episodes.  As {expected}, the \ac{UAV} named ``A", whose position was closer to the target, got sooner to the destination position but with a scarce map reconstruction. By contrast, the \ac{UAV} ``B" was capable of reconstructing a better copy of the map, and of finding a good trajectory after some training episodes thanks to the aid offered by the other \ac{UAV}.

Note also that both agents were not efficient during the first episode (Fig.~\ref{fig:fig7}-(b,c)). This implies that if the task has to be accomplished only in one episode, the swarm might not correctly accomplish the goal of the mission. 

Finally, Fig.~\ref{fig:fig8} shows the behavior of the positive $Q$-values (e.g., positive rewards received from the environment) as a function of the number of episodes and averaged over ten simulations. Notably, when multiple \acp{UAV} update a common $Q$-table, the time needed in terms of episodes to achieve a greater amount of positive rewards (i.e., a better knowledge of the environment) is less than that required by a single agent.
This result confirms that the employment of many coordinated \acp{UAV} allows to accelerate the learning procedure and, thus, to enable the exploration of large environments. On the other side, when the network is composed of many \acp{UAV}, the consequent communication burden with the edge should be accounted for. 

\section{Conclusions \& Future Perspectives}
\label{sec:future}

It is clear that \acp{UAV} present a wide range of requirements: they need to be autonomous, fast, multi-tasking, and intelligent. But, above all, they should preserve a low complexity {implementation}  while accurately accomplishing the mission for which they have been deployed. 
Consequently, the on-board \ac{UAV-I} needs to be properly designed and, when necessary, delegated to a higher entity of the network thanks to a communication scheme enabling such distribution. 
In this sense, even if the performance of the case studies depends on the specific set-up, some general conclusions can be drawn. % and a summary is provided in the following.

\subsubsection{Final Remarks on Communication}
In {\emph{Use Case 1}}, we compared a case in which \acp{UAV} lean on an edge and one with multi-hops. In the edge scenario, the availability of a global view of the environment naturally improves the accuracy of the estimate, but the performance depends on the latency due to access and processing delays. Indeed, the presence of aged measurements might lead to an old view of the state of the system, especially in fast-changing environments. 

In the case with multi--hops, the delays are dictated by the connections created at each time instant between the nodes. Therefore, if the network is small and can be connected with a single-hop, it allows for processing of all sensed data, as for an edge-aided network, with the advantage of being less vulnerable. The downside is that some \acp{UAV} could be isolated without receiving any information from the other \acp{UAV}, creating very high localization errors and inducing higher latency in finding a good flying formation.

\subsubsection{Final Remarks on Navigation}

In this paper we have provided two trajectory design schemes for the \ac{UAV} network, a gradient-based and a learning-based method. The first approach is certainly faster and allows, after a few instants, to obtain the desired \ac{UAV} formation. By contrast,  the second approach requires a longer learning time (e.g., different training episodes). The first approach is good when there is a model of the function that we want to optimize but is subject to modeling errors. For example, based on the chosen communication scheme, the model could be affected by delayed information (e.g., coming from multi-hop measurements) or inaccurate position estimates. The second approach is more robust to model mismatches and can be accelerated thanks to collaboration {among} the \acp{UAV} or with the edge.
\subsubsection{Final Remarks on Complexity}  With {\emph{Use Case 1}} we have evidenced that, even for a low-dimensional state, the edge can help in reducing the time required for finding the best swarm formation for getting closer to a target. This aspect can be crucial when the available time is limited and, thus, the off-load of computations might be for low-dimensional states. Note that the application itself drives the design of the \ac{UAV-I}: for example, relying on edges for mission--critical applications could be fatal due to network vulnerability to external attacks. In fact, if the edge is hacked by an external user, the entire network will not be capable to perform the assigned task in safe conditions. Moreover, if the \acp{UAV} are used in post-disaster situations, the edges (e.g., the cellular infrastructure) might not be accessible. Also, to further reduce delays, it is recommended not to delegate anti-collision operations to external nodes.

Similarly, in \emph{Use Case 2}, we highlighted that dealing with a large $Q$–table might be complicated for low-complexity \acp{UAV}. Such operations should be delegated to an edge, with consequent privacy and security issues.

\subsubsection{Final Remarks on Network Scalability} Because time is a critical resource, the acceleration of the learning procedure is an intrinsic task for future swarms of \acp{UAV}. In this sense, apart from the adoption of more performing learning algorithms, cooperation and coordination techniques for \acp{UAV} can be of great help for speeding up the process, especially in scenarios where the rewards are sparse in the environment.
On the other side, this  implies an increase of hardware complexity and of coordination among \acp{UAV} which have to interact {between} each other and eventually with the edge. 
%Thus, finding the trade-off between performance and scalability is still an open research issue, as the optimal number of employed agents has to guarantee reliable performance and an overall affordable complexity.

\subsubsection{Future Perspectives}
In future autonomous networks, %the 
teams of \acp{UAV} are expected to be capable to self-evaluate the goodness of their performance through a measure of proficiency. Such metric, together with received environmental feedback, will provide an important assessment to each \ac{UAV} to optimize the performance and re-assess each time when new observations are acquired. %Thus, the \acp{UAV} will need to be aware of the effort required for each  activity to be completed based on their experience and to understand when to delegate them to other parts of the network.

A step ahead of the self-assessment is represented by situation-awareness, which %that is, to 
accounts for information about the context in which the \acp{UAV} accomplish their tasks. In this direction, one possibility is to dynamically adapt the communication scheme according to the needs and conditions experienced during the \ac{UAV} flight time.
%We refer to this scheme as beyond \ac{U2U} communication \cite{zhang2020beyond}. 
In this light, future generations of \ac{UAV} networks will exploit the \ac{UAV-I} to control {the communication scheme} by alternating the destinations based on different time-varying requirements (e.g., offloading the computation or preserving low latency).
%\medskip

%\textbf{UAV accelerated-learning}: 
%In our \emph{Use Case 2} we have provided an example of learning-based application without analyzing the impact of the different learning approaches into the attainable performance. Again, as in \cite{lee2020optimization}, an extensive analysis of possible approaches has already been provided. Nevertheless, since time is a critical resource, the acceleration of the learning procedure is a mandatory task for future swarms of \acp{UAV} employed for emergency situations where localization is required. In this sense, the exploitation of hundreds of coordinated \acp{UAV} can be of great help in speed-up the learning process.
%\medskip

Finally, in our case study we have experienced that the independent design of sensing and communication might not be flexible enough to simultaneously satisfy the needs of the application because the control dictates the most stringent communication requirements to attain some sensing performance but not vice-versa. From this point of view, a co-design is preferable to jointly maximize a twofold communication and sensing performance, even if it may be particularly challenging as their requirements often contrast with each other. In this sense, further coexistence  problems arise when they simultaneously take place in the same frequency bands. For example, \ac{ML} can be adopted for picking the best \ac{UAV} trajectories while optimizing the sensing and communication parameters.
%, or they can act as classifiers in radar operations to perform an accurate target classification in complex environments. 
%Moreover, for the specific case of radar sensors, the co-design tackles the possibility to share the same hardware for carrying out both functionalities through the same radio waveform. %{\color{red} DD: not clear. This statement is true only for the specific case of radar sensors. }.


\end{document}